\documentclass[12pt,epsf]{article}
\usepackage{graphicx}
\begin{document}


\def\a{\alpha}
\def\b{\beta}
\def\c{\varepsilon}
\def\d{\delta}
\def\e{\epsilon}
\def\f{\phi}
\def\g{\gamma}
\def\h{\theta}
\def\k{\kappa}
\def\l{\lambda}
\def\m{\mu}
\def\n{\nu}
\def\p{\psi}
\def\q{\partial}
\def\r{\rho}
\def\s{\sigma}
\def\t{\tau}
\def\u{\upsilon}
\def\v{\varphi}
\def\w{\omega}
\def\x{\xi}
\def\y{\eta}
\def\z{\zeta}
\def\D{\Delta}
\def\G{\Gamma}
\def\H{\Theta}
\def\L{\Lambda}
\def\F{\Phi}
\def\P{\Psi}
\def\S{\Sigma}

\def\o{\over}
\def\beq{\begin{eqnarray}}
\def\eeq{\end{eqnarray}}
\newcommand{\gsim}{ \mathop{}_{\textstyle \sim}^{\textstyle >} }
\newcommand{\lsim}{ \mathop{}_{\textstyle \sim}^{\textstyle <} }
\newcommand{\vev}[1]{ \left\langle {#1} \right\rangle }
\newcommand{\bra}[1]{ \langle {#1} | }
\newcommand{\ket}[1]{ | {#1} \rangle }
\newcommand{\EV}{ {\rm eV} }
\newcommand{\KEV}{ {\rm keV} }
\newcommand{\MEV}{ {\rm MeV} }
\newcommand{\GEV}{ {\rm GeV} }
\newcommand{\TEV}{ {\rm TeV} }
\def\diag{\mathop{\rm diag}\nolimits}
\def\Spin{\mathop{\rm Spin}}
\def\SO{\mathop{\rm SO}}
\def\O{\mathop{\rm O}}
\def\SU{\mathop{\rm SU}}
\def\U{\mathop{\rm U}}
\def\Sp{\mathop{\rm Sp}}
\def\SL{\mathop{\rm SL}}
\def\tr{\mathop{\rm tr}}

\def\IJMP{Int.~J.~Mod.~Phys. }
\def\MPL{Mod.~Phys.~Lett. }
\def\NP{Nucl.~Phys. }
\def\PL{Phys.~Lett. }
\def\PR{Phys.~Rev. }
\def\PRL{Phys.~Rev.~Lett. }
\def\PTP{Prog.~Theor.~Phys. }
\def\ZP{Z.~Phys. }

\def\stau{\widetilde{\tau}}
\def\mstau{m_{\stau}}
\def\gravitino{\widetilde{G}}
\def\mgravitino{m_{\gravitino}}
\def\Mpl{M_\mathrm{P}}


\baselineskip 0.7cm

\begin{titlepage}

\begin{flushright}
  DESY-064
  \\
  UT-06-05
\end{flushright}

\vskip 1.35cm
\begin{center}
{\large \bf
Eluding the BBN constraints on the stable gravitino
}
\vskip 1.2cm
W. Buchm\"uller$^1$, K. Hamaguchi$^{1,2}$,
M. Ibe$^2$, T. T. Yanagida$^{1,2}$
\vskip 0.4cm

${}^1${\it Deutsches Elektronen-Synchrotron DESY, 22603 Hamburg, Germany}\\
${}^2${\it Department of Physics, University of Tokyo,
     Tokyo 113-0033, Japan}

\vskip 1.5cm

\abstract{ We investigate how late-time entropy production weakens the
Big-Bang Nucleosynthesis (BBN) constraints on the gravitino as
lightest superparticle with a charged slepton as next-to-lightest
superparticle. We find that with a moderate amount of entropy
production, the BBN constraints can be eluded for most of the
parameter space relevant for the discovery of the gravitino. This is
encouraging for experimental tests of supergravity at LHC and ILC. }
\end{center}
\end{titlepage}

\setcounter{page}{2}

\noindent 
{\it Introduction}

The gravitino $\gravitino$ is a unique and inevitable prediction of
supergravity (SUGRA)~\cite{SUGRA}, and hence the discovery of the
gravitino would provide unequivocal evidence for SUGRA. It has been
pointed out that this test of SUGRA may be possible at LHC or ILC, if
the gravitino is the lightest superparticle (LSP) and the long-lived
next-to-lightest superparticle (NLSP) is a charged
slepton~\cite{Buchmuller:2004rq}.

{}From an experimental point of view, a relatively large gravitino
mass $\mgravitino$ comparable to the slepton mass $m_{\tilde{\ell}}$,
$\mgravitino \gsim {\cal O}(0.1)\, m_{\tilde{\ell}}$, is particularly
interesting~\cite{Buchmuller:2004rq,Hamaguchi:2004df+X}.  This is
because in such a gravitino mass region the kinematical reconstruction
of the gravitino mass becomes possible, which leads to a determination
of the ``Planck scale'', and even the gravitino spin might become
measurable.

However, such a parameter region is strongly constrained by cosmology.
In particular, the BBN constraints on a late decaying
particle~\cite{Kawasaki:2004qu,Jedamzik:2006xz} lead to an upper bound
on the gravitino mass for a given slepton
mass~\cite{Asaka:2000zh+Fujii:2003nr,Feng:2004mt}, which makes the
SUGRA test at collider experiments very challenging.

It is, however, easy to evade the BBN constraints if late-time entropy
production occurs after the slepton decoupling (and before BBN). In
this letter we explicitly show how much late-time entropy production
weakens the BBN constraints on the NLSP decay into the gravitino. We
find most of the relevant parameter space to survive for a moderate
amount of entropy production.  This is very encouraging with respect
to experimental tests of SUGRA at LHC and ILC. It has also interesting
implications for leptogenesis, which will be discussed
elsewhere~\cite{BHIY2}.
\vspace{.6cm}

\noindent
{\it BBN constraint with late-time entropy production}

For concreteness, we assume that the NLSP is the superpartner of the
tau lepton, stau ($\stau$). In the early universe, the stau NLSP is in
thermal equilibrium until its decoupling at $T_d\sim \mstau/20$. If
the stau particle decays during or after BBN, $T_{BBN}\sim
1~\mathrm{MeV}$, it may spoil the successful BBN
predictions~\cite{Kawasaki:2004qu,Jedamzik:2006xz}. In the model with
stau NLSP and gravitino LSP, this leads to severe constraints on the
parameter space $(\mstau,\mgravitino)$, in particular to upper bounds
on the gravitino mass for a given stau
mass~\cite{Asaka:2000zh+Fujii:2003nr,Feng:2004mt}.

If there is no entropy production after the stau decoupling, the thermal relic
abundance of the stau before its decay is given
by~\cite{Asaka:2000zh+Fujii:2003nr}
\begin{eqnarray}
  Y_{\stau}^{\mathrm{thermal}}
  \;\equiv\; 
  \frac{n_{\stau}}{s} 
  \;=\; 
  \kappa \times 10^{-13} \left(\frac{\mstau}{100~\mathrm{GeV}}\right),
 \label{eq:Ystau}
\end{eqnarray}
where $n_{\stau}$ and $s$ are the stau number density and the
entropy density, respectively.
Here, $\kappa=(0.7-1)$ is a numerical coefficient which depends on the model parameters (e.g.,
$\tan\beta$). 
In the following we take $\kappa=0.7$ as a representative value. 
The decay rate of the stau NLSP is given
by~\cite{Buchmuller:2004rq}
\begin{eqnarray}
  \Gamma_{\stau}(\stau\to\gravitino\tau)
  &=&
  \frac{\mstau^5}
       {48\pi \mgravitino^2 \Mpl^2}
       \left(1-
       \frac{\mgravitino^2 + m_\tau^2}
	    {\mstau^2}
	    \right)^4
	    \left[
	    1-\frac{4 \mgravitino^2 m_\tau^2}
	    {(\mstau^2-\mgravitino^2-m_\tau^2)^2}
	    \right]^{3/2},
	    \nonumber\\
	    &\simeq&
	    (6\times 10^6~\mathrm{sec})^{-1}
	    \left(\frac{\mstau}{100~\mathrm{GeV}}\right)^5
	    \left(\frac{10~\mathrm{GeV}}{\mgravitino}\right)^2
	    \left(1-
	    \frac{\mgravitino^2}{\mstau^2}
	    \right)^4\;,
\end{eqnarray}
where in the second equation we have neglected the mass of the
tau--lepton, $m_{\t}$.

The energetic tau-lepton produced by the stau decay causes the
electromagnetic (EM) cascade, which results in destructions or
overproductions of light elements (D, ${}^{3}$He, ${}^{4}$He, etc.).
However, the tau-lepton itself decays before interacting with
background photons, and hence, some of the energy carried by the
tau-lepton is lost to neutrinos.  Thus, the electromagnetic energy
released by the stau decay is given by
\begin{eqnarray}
  E_{EM} Y_{\stau} = \xi_{EM} \frac{\mstau^2-\mgravitino^2}{2 \mstau} Y_{\stau},
\end{eqnarray}
where $\xi_{EM}\lsim 1$ denotes the suppression factor due to the
energy loss in the tau decay into neutrinos.  In the following, we
take $\xi_{EM}\simeq 0.5$ as a representative
value~\cite{Feng:2004mt}.  On the other hand, the hadronic
contribution relevant for BBN dominantly comes from the three- and
four-body decay of the stau, and therefore the hadronic branching
ratio of the stau NLSP is suppressed, $B_{h}\lsim
10^{-3}$~\cite{Feng:2004zu}.  Thus, the hadronic energy released by
the stau decay is suppressed by a factor $\lsim 10^{-3}$ compared to
the EM energy.

Without late-time entropy production, the parameter region with a
relatively large gravitino mass, $\mgravitino \gsim {\cal O}(0.1)\,
m_{\stau}$, which is wanted for tests of SUGRA, is severely
constrained by the above BBN constraints.  However, if an adequate
entropy production occurs after the stau decoupling, the stau
abundance is diluted by a factor $\Delta$,
\begin{eqnarray}
  Y_{\stau} = \frac{1}{\Delta} Y_{\stau}^{\mathrm{thermal}},
\end{eqnarray}
and the BBN constraints can be easily eluded.%
\footnote{If the entropy is provided by the late-time decay of a
long-lived particle $\varphi$, the dilution factor $\D$ is given by
$\D \simeq ( T_{d}/ T_{\varphi} )^{3}$ in terms of the decay
temperature of $\varphi$, $T_{\varphi}$ (assuming that the direct
production of $\stau$ from $\varphi$ is negligible). Thus, for
instance, the dilution factor is $\D\sim 10^3$ for $T_\varphi\sim
1$~GeV and $\mstau\simeq 200$~GeV.}

\begin{figure}[t!]
  \begin{center}
	\includegraphics[width=.5\linewidth]{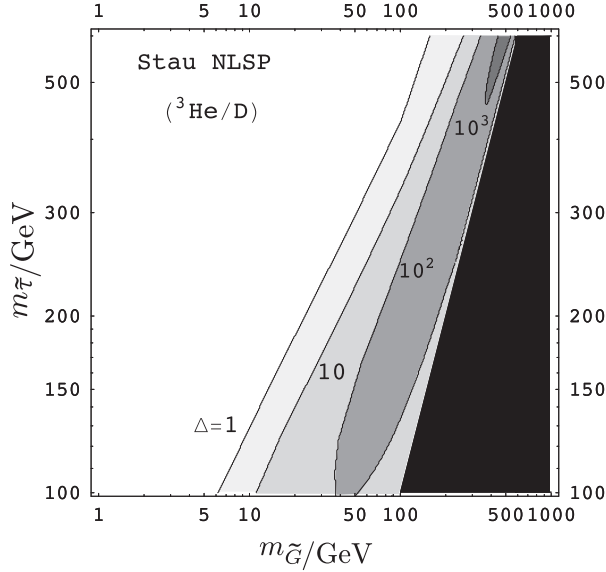} 
  \end{center}
  \caption{The BBN constraint (${}^{3}$He/D bound) on the parameter space 
  $( \mgravitino, \mstau)$ with late-time entropy production.
 The regions excluded by the ${}^{3}$He/D bound
 are shaded from light to dark gray 
 for a dilution factor $\Delta=1$, 10, $10^{2}$, $10^{3}$.
 In the black shaded region, the gravitino is not the LSP.
  Here we have neglected the effects of hadronic decay (see text). }
  \label{fig:mstau_mgra}
\end{figure}

To show this explicitly, we plot the BBN constraints in the
$(\mgravitino, \mstau)$ plane (Fig.~\ref{fig:mstau_mgra}).  Here, we
have used the constraint from the ${}^{3}$He/D bound, and neglected
other photo-dissociation and hadro-dissociation effects on the light
elements.  This is because, in the region $m_{\stau} Y_{\stau} \lsim
10^{-11}$~GeV, the most stringent constraint comes from the ${}^{3}$He
overproduction for a late decaying particle with $B_{h}\lsim 10^{-3}$
(see Fig.~41 and Fig.~42 in Ref.~\cite{Kawasaki:2004qu}).  Thus, for
our purposes, the ${}^{3}$He/D bound is the most stringent BBN
constraint for $m_{\stau}\lsim 500$\,GeV and $\Delta\gsim 10$.%
\footnote{In Fig.~\ref{fig:mstau_mgra}, we have used the constraints
in Fig.~41 of Ref.~\cite{Kawasaki:2004qu}. Our $\D = 1$ line in
Fig.~\ref{fig:mstau_mgra} almost reproduces the ${}^{3}$He/D bound in
Fig.~3 of Ref.~\cite{Feng:2004mt}.}

As Fig.~\ref{fig:mstau_mgra} demonstrates, the ${}^{3}$He/D bound
severely constrains the parameter space for $\D = 1$, while the most of
the interesting region is allowed for $\D=10^{3}$.  Therefore, the
bulk of the relevant parameter space for the SUGRA test survives with
a moderate amount of entropy production. Consequences of the present
analysis for leptogenesis will be discussed in Ref.~\cite{BHIY2}.

\section*{Acknowledgments}
M.~I. thanks the Japan Society for the Promotion of Science for
financial support.  This work is partially supported by Grand-in-Aid
Scientific Research (s) 14102004.
The work of T.T.Y. has been supported in part by a Humboldt Research Award.


\begin{thebibliography}{99}
%

\bibitem{SUGRA}
  D.~Z.~Freedman, P.~van Nieuwenhuizen and S.~Ferrara,
  Phys.\ Rev.\ D {\bf 13} (1976) 3214;
  \\
  S.~Deser and B.~Zumino,
  Phys.\ Lett.\ B {\bf 62} (1976) 335.


\bibitem{Buchmuller:2004rq}
  W.~Buchmuller, K.~Hamaguchi, M.~Ratz and T.~Yanagida,
  Phys.\ Lett.\ B {\bf 588} (2004) 90
  [arXiv:hep-ph/0402179];
  %
  arXiv:hep-ph/0403203.


\bibitem{Hamaguchi:2004df+X}
  K.~Hamaguchi, Y.~Kuno, T.~Nakaya and M.~M.~Nojiri,
  Phys.\ Rev.\ D {\bf 70} (2004) 115007
  [arXiv:hep-ph/0409248];\\
  J.~L.~Feng and B.~T.~Smith,
  Phys.\ Rev.\ D {\bf 71}, 015004 (2005)
  [Erratum-ibid.\ D {\bf 71}, 0109904 (2005)]
  [arXiv:hep-ph/0409278];\\
  K.~Hamaguchi, M.~M.~Nojiri, A.~de Roeck, in preparation.


\bibitem{Kawasaki:2004qu}
  M.~Kawasaki, K.~Kohri and T.~Moroi,
  Phys.\ Rev.\ D {\bf 71} (2005) 083502
  [arXiv:astro-ph/0408426],
  and references therein.



\bibitem{Jedamzik:2006xz}
  K.~Jedamzik,
  arXiv:hep-ph/0604251,
  and references therein.


\bibitem{Asaka:2000zh+Fujii:2003nr}
  T.~Asaka, K.~Hamaguchi and K.~Suzuki,
  Phys.\ Lett.\ B {\bf 490} (2000) 136
  [arXiv:hep-ph/0005136];
  \\
  M.~Fujii, M.~Ibe and T.~Yanagida,
  Phys.\ Lett.\ B {\bf 579} (2004) 6
  [arXiv:hep-ph/0310142].


\bibitem{Feng:2004mt}
  J.~L.~Feng, S.~Su and F.~Takayama,
  Phys.\ Rev.\ D {\bf 70} (2004) 075019
  [arXiv:hep-ph/0404231].

\bibitem{BHIY2}
W.~Buchmuller, K.~Hamaguchi, M.~Ibe, and T.~T.~Yanagida, in preparation.


\bibitem{Feng:2004zu}
  J.~L.~Feng, S.~Su and F.~Takayama,
  Phys.\ Rev.\ D {\bf 70}, 063514 (2004)
  [arXiv:hep-ph/0404198].



\end{thebibliography}
\end{document}